\documentclass[a4paper]{iopart}

\usepackage[english]{babel}
\usepackage[utf8x]{inputenc}

\usepackage{textcomp,marvosym}
\usepackage{cite}
\usepackage{color}
\usepackage{amssymb}
\usepackage{multirow}
\usepackage{graphicx}
\usepackage{paralist}
\usepackage[colorinlistoftodos]{todonotes}
\usepackage[colorlinks=true, allcolors=blue]{hyperref}

\newcommand{\aligo}{Advanced LIGO}
\newcommand{\pycbc}{\texttt{PyCBC}}

\begin{document}

\title[]{Blip glitches in \aligo{} data}

\author{M. Cabero$^{1,2}$, A. Lundgren$^{1,3}$, A. H. Nitz$^1$, T. Dent$^{1,4}$, D. Barker$^5$, E. Goetz$^5$, J. S. Kissel$^5$, L. K. Nuttall$^{3,6}$, P. Schale$^7$, R. Schofield$^{5,7}$ and D. Davis$^8$}
\address{$^1$ Max Planck Institute for Gravitational Physics (Albert Einstein Institute), Callinstrasse 38, D-30167 Hannover, Germany}
\address{$^2$ Department of Physics, Princeton University, Princeton, NJ 08544, USA}
\address{$^3$ Institute of Cosmology and Gravitation, University of Portsmouth, Dennis Sciama Building, Burnaby Road, Portsmouth, PO1 3FX, United Kingdom}
\address{$^4$ Instituto Galego de Física de Altas Enerxías,
Universidade de Santiago de Compostela, E-15782 Santiago de Compostela, Spain}
\address{$^5$ LIGO Hanford Observatory, P.O. Box 159, Richland, WA 99352, USA}
\address{$^6$ School of Physics and Astronomy, Cardiff University, Cardiff, CF24 3AA, United Kingdom}
\address{$^7$ Department of Physics, University of Oregon, Eugene, OR 97403, USA}
\address{$^8$ Department of Physics, Syracuse University, Syracuse, NY 13244, USA}

\ead{mcmuller@princeton.edu}

\date{\today}

\begin{abstract}
Blip glitches are short noise transients present in data from ground-based gravitational-wave observatories. These glitches resemble the gravitational-wave signature of massive binary black hole mergers. Hence, the sensitivity of transient gravitational-wave searches to such high-mass systems and other potential short duration sources is degraded by the presence of blip glitches. The origin and rate of occurrence of this type of glitch have been largely unknown. In this paper we explore the population of blip glitches in \aligo{} during its first and second observing runs. On average, we find that \aligo{} data contains approximately two blip glitches per hour of data. We identify four subsets of blip glitches correlated with detector auxiliary or environmental sensor channels, however the physical causes of the majority of blips remain unclear.
\end{abstract}

\maketitle

\section{Introduction}
\label{sec:introduction}

The Laser Interferometer Gravitational-wave Observatories (LIGO)~\cite{AdvLIGO} 
and the Virgo observatory~\cite{AdvVirgo} have successfully identified several 
mergers of compact binaries in their second generation (Advanced) configuration.
The first observing run of \aligo{} (O1) took place from September 12{$^{\rm th}$} 2015 
to January 19{$^{\rm th}$} 2016, concluding with the observation of three binary black hole 
coalescences~\cite{GW150914,GW151226,O1BBH}. The second observing run (O2) started in 
November 30{$^{\rm th}$} 2016 and lasted until August 25{$^{\rm th}$} 2017, 
with Advanced Virgo joining the second generation network in August 2017. 
The O2 run brought the first gravitational wave of a coalescing 
neutron star~\cite{GW170817}, as well as several other black-hole 
mergers~\cite{GW170104,GW170608,GW170814,O2catalog}.

The sensitivity of interferometric gravitational-wave detectors to astrophysical signals may be
assessed via the power spectral density (PSD) of a stationary Gaussian process describing noise 
contributions to the measured strain~\cite{Martynov:2016fzi}. 
This description is commonly used in assessing the long-term evolution of detector performance. 
However, it neglects the presence of transient (short duration) non-Gaussian features in 
the detector noise~\cite{Blackburn:2008ah,TheLIGOScientific:2016zmo}, known as `glitches'. 
The presence of noise transients can substantially degrade the sensitivity of
searches for transient gravitational-wave signals in the strain data~\cite{O1_DQ}.  
Thus, it is of interest to describe, understand and, if possible, mitigate glitches.

A particular type of short duration noise transient commonly known as
`blip glitch'~\cite{TheLIGOScientific:2016zmo, O1_DQ, O2catalog}
is amongst the worst contributors to the background of transient
gravitational-wave searches. Determining the source of blip glitches 
is crucial to improve the sensitivity of these searches. However,
there are many different types of noise transients in the data and
computer algorithms cannot easily distinguish blip glitches from
other short-impulse noise transients. 
Here, we develop a method to identify blip glitches in gravitational-wave
data and study possible sources of these noise transients in the Advanced
LIGO detectors.

This manuscript is organised as follows. Sections~\ref{sec:detectors} provides an overview
on the Advanced LIGO detectors. Section~\ref{sec:definition} describes the morphology 
and characteristics of a blip glitch. In Sec.~\ref{sec:finding_blips} we develop a method to 
identify times of blip glitches using \pycbc{} 
tools~\cite{PyCBCgithub, Canton:2014ena, Usman:2015kfa}. 
In Sec.~\ref{sec:investigating_blips} we investigate the origin of these blip glitches 
identified with \pycbc{}. Finally, the manuscript is summarised in Sec.~\ref{sec:conclusions}.


\section{The Advanced LIGO detectors}
\label{sec:detectors}

\begin{figure}[tb]
\centering
\includegraphics[width=0.9\columnwidth]{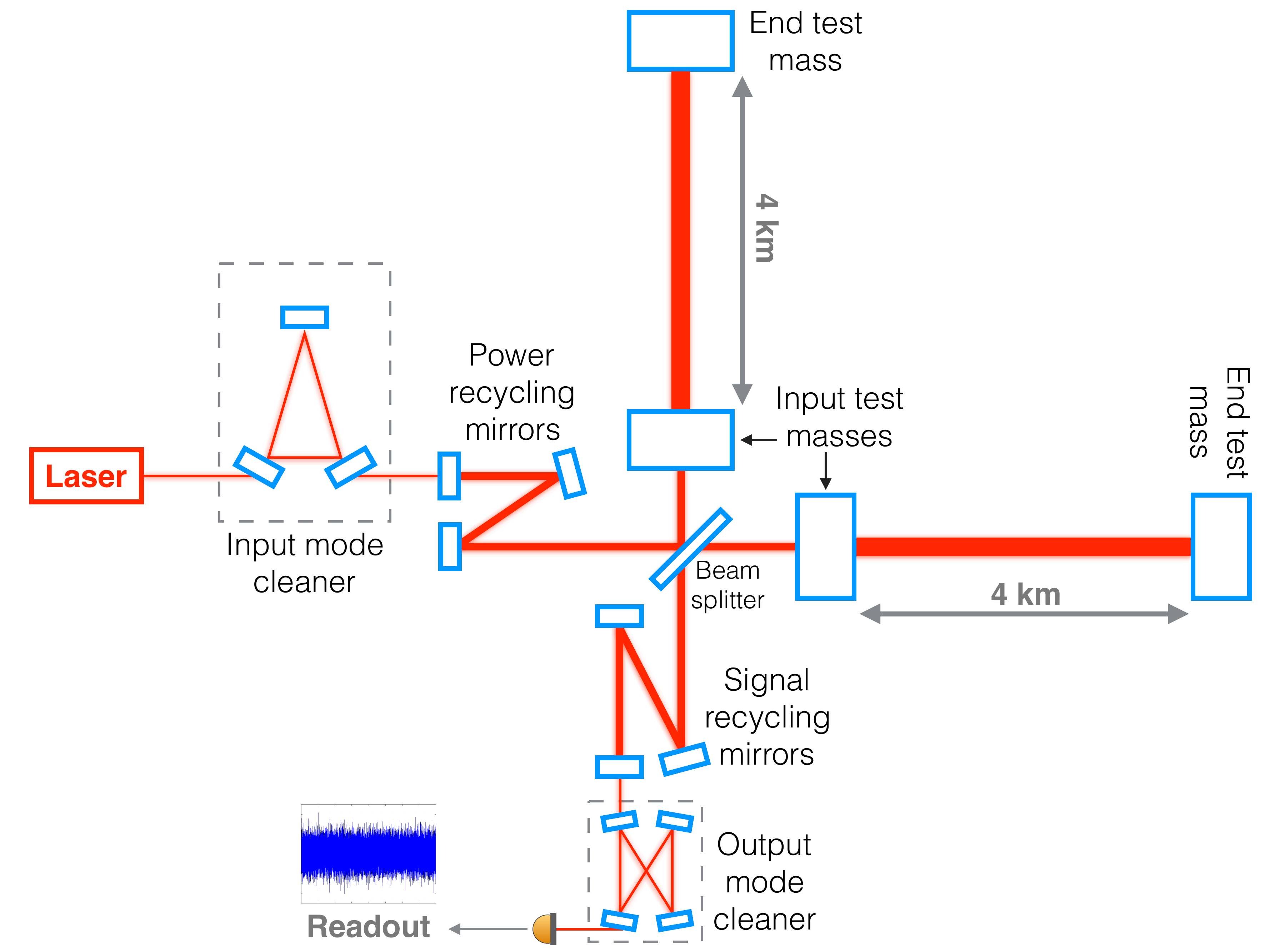}
\caption{\label{fig:advLIGOconfig} Simplified optical configuration of the \aligo{} detectors (original image in~\cite{Cabero:thesis}).}
\end{figure}

The basic design of the Advanced LIGO gravitational-wave 
detectors~\cite{AdvLIGO,PhysRevLett.116.131103} 
is a Michelson interferometer with Fabry-P\'{e}rot resonant cavities 
in each of the 4-km long arms. 
The resonance condition of the optical cavities is maintained by multiple 
servo-control loops.
A simplified sketch of the optical
configuration of an \aligo{} interferometer is shown in 
Fig.~\ref{fig:advLIGOconfig}. The input mode cleaner stabilises the 
laser frequency and suppresses higher order spatial modes before the 
light enters the interferometer~\cite{doi:10.1063/1.4936974}.
Between the input mode cleaner and the beam splitter, 
the power recycling mirror is placed to increase the effective laser power. 
The input and end test masses of the Fabry-P\'{e}rot arm cavities are 
suspended by a quadruple pendulum system~\cite{Aston:2012ona, Carbone:2012bd}
and serve as test masses. 
The quadruple suspensions are mounted to actively stabilised in-vacuum 
optical tables that provide seismic isolation from the environment~\cite{Matichard:2015eva}.
At the anti-symmetric output of the Michelson, the signal recycling mirror 
is used to maintain a broad frequency response of the detector. 
Finally, the output mode cleaner~\cite{0264-9381-29-6-065005} filters out
higher order spatial modes produced in the interferometer before the light
enters the readout photodiodes. The signal measured by the photodetectors 
is digitised and calibrated to convert laser light power to relative mirror
displacement~\cite{PhysRevD.95.062003, Viets:2017yvy, Acernese:2018bfl}. 
This calibrated strain signal, $h(t)$, is the data analysed in the searches for gravitational waves.

The detector's strain sensitivity to astrophysical signals is mainly
limited by fundamental noise sources~\cite{AdvLIGO}.
Thermal noise, stronger at lower frequencies, arises from the test masses 
and their suspension systems. Quantum noise includes shot noise at high 
frequencies and radiation pressure at low frequencies. Seismic noise is the 
main limiting factor at low frequencies. 
Technical noise sources, which are controllable by design, further
shape the detector's sensitivity. Several major upgrades were implemented
between the Initial and Advanced LIGO generations to reduce these noise
sources and improve the detectors' sensitivities. For instance, 
the signal recycling cavity and the new quadruple pendulum suspension
mentioned above were first introduced in the Advanced LIGO era. 
Moreover, larger and heavier test masses were built to reduce 
thermal noise and motion induced by quantum radiation pressure.
During the first observing run of Advanced LIGO, the detectors' strain 
noise was already between 4 and 30 times lower than in the final science 
run of Initial LIGO~\cite{PhysRevLett.116.131103},
depending on the frequency band. 
Further improvements will be implemented in the coming years until Advanced 
LIGO's design sensitivity is reached, such as the gradual increase of the 
laser power and the injection of squeezed light to reduce quantum 
noise~\cite{AdvLIGO, PhysRevLett.116.131103, Miller:2014kma}.

Technical, or hardware, and environmental noise sources can produce glitches visible 
in the calibrated strain data that limit the search's sensitivity. 
Identifying the source of a particular glitch in a gravitational-wave detector is 
very challenging: in addition to the main gravitational-wave channel, there are over
200,000 auxiliary channels in each detector.
These channels monitor the environmental conditions around the detector and 
the hardware behaviour of the interferometer~\cite{0264-9381-32-3-035017}. 
Should a source for a certain type of glitch be identified, the detector issue may be 
fixed at the source to mitigate such glitches~\cite{Nuttall:2015dqa, Nuttall:2018xhi, Berger:2018ckp}. 
Alternatively, albeit less effectively, auxiliary channels can be used to identify 
times of occurrences and discard -- referred to as a veto -- the glitches from the
$h(t)$ data used in gravitational-wave 
searches~\cite{TheLIGOScientific:2016zmo, O1_DQ, McIver:2012ky}. 
Unfortunately, blip glitches are still of unknown origin and cannot be mitigated or safely vetoed.


\section{Description of a blip glitch}
\label{sec:definition}

\begin{figure}[b]
\centering
\includegraphics[width=0.45\columnwidth]{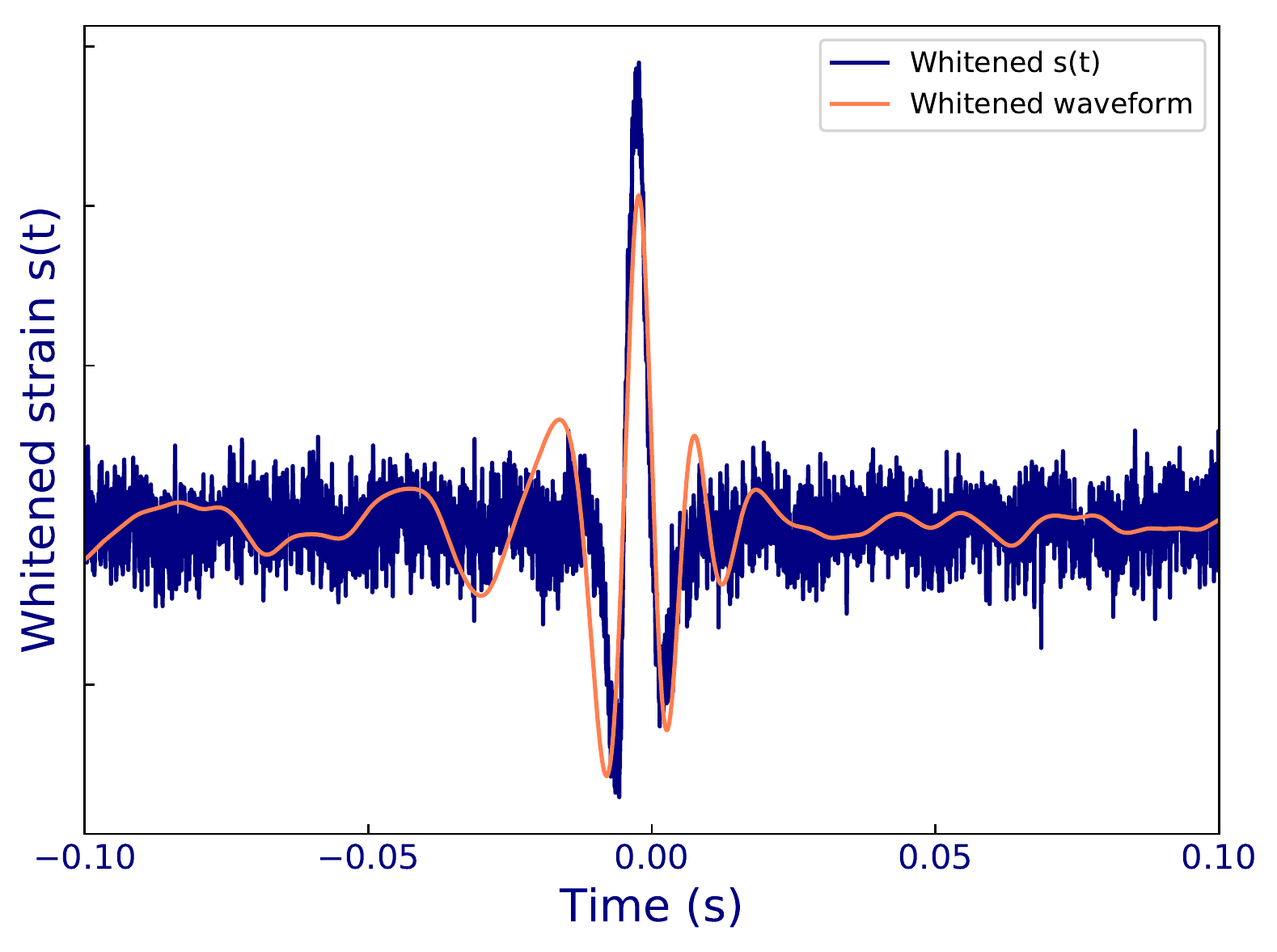}
\includegraphics[width=0.45\columnwidth]{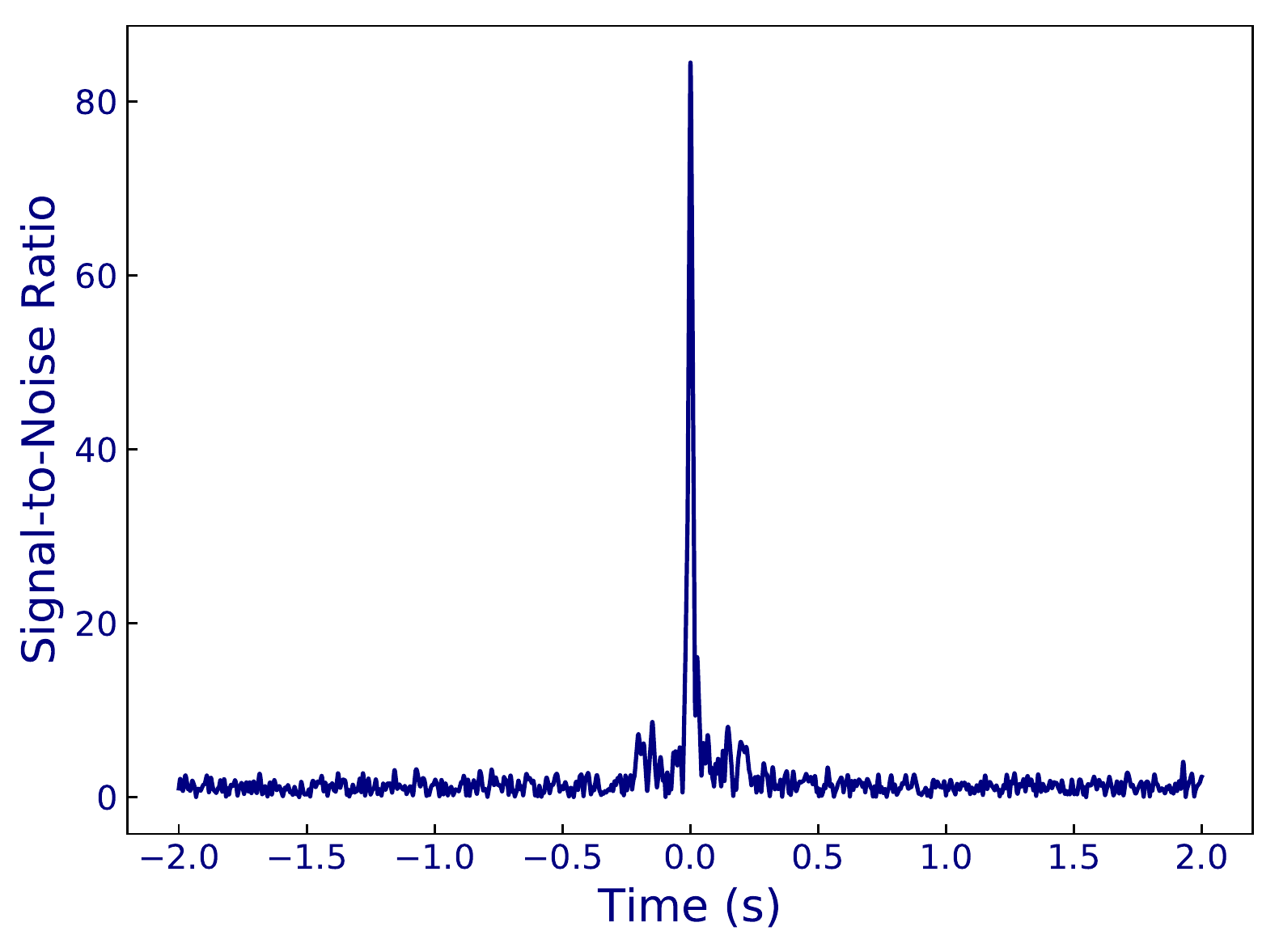}
\caption[]{\label{fig:Timeseries_blips} (\textit{Left}) Whitened $h(t)$ strain at the time of a blip glitch with a whitened compact binary waveform overlaid on top of it. The waveform corresponds to a system with a high total mass ($M_t = 95.7 M_\odot$) and a large mass ratio ($q=m_1/m_2\simeq22$). (\textit{Right}) Signal-to-noise ratio obtained by matched filtering the template and the data shown on the left plot (original image in~\cite{Cabero:thesis}).}
\end{figure}

A noise transient is categorised as a blip glitch if it is a very short duration
transient, $\mathcal{O}(10)$ ms, with a large frequency bandwidth, $\mathcal{O}(100)$ Hz.
Blip glitches are found in both LIGO detectors, located in Hanford (Washington state)
and Livingston (Louisiana). There is also evidence for the presence of similar short
noise transients in the Virgo (Italy) and GEO~600 (Germany) gravitational-wave 
observatories. We focus here on blip glitches observed in \aligo{} data.

The characteristic time-domain shape of a blip glitch resembles the 
gravitational-wave imprint of compact binaries with large total mass, highly asymmetric
component masses and spins anti-aligned with the orbital angular momentum 
(see Fig.~\ref{fig:Timeseries_blips}).
While occurrences of blip glitches are independent between the two LIGO interferometers,
blip glitches contribute to the background of transient gravitational-wave searches.
This is because there is a nonzero probability of random (accidental) coincidences in 
time between blip glitches in different detectors, or between blips and random noise; 
this probability is estimated by performing analyses with unphysical time-shifts 
between detectors~\cite{Usman:2015kfa}. Hence, the presence of blip glitches degrades the search sensitivity
to high-mass systems.

\begin{figure}[tb]
\centering
\includegraphics[width=0.6\columnwidth]{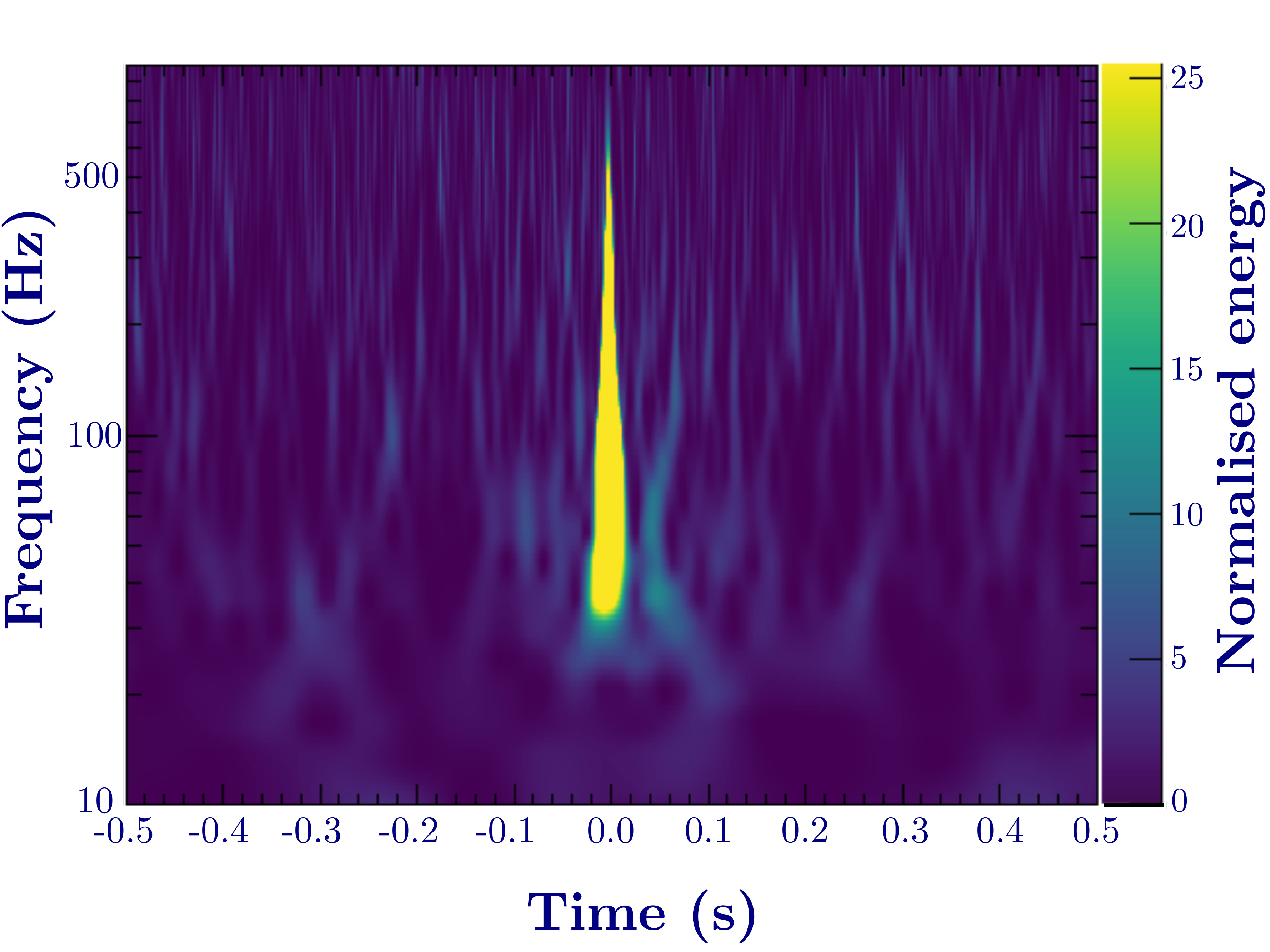}
\caption[]{\label{fig:Blip_glitch} Time-frequency representation of a typical blip glitch, seen in Hanford data on August 21st 2017 (original image in~\cite{Cabero:thesis}).}
\end{figure}

A time-frequency representation~\cite{0264-9381-21-20-024, Qtransform} of a typical
blip glitch is shown in Fig.~\ref{fig:Blip_glitch}. This glitch morphology in the
calibrated gravitational-wave channel was first observed in Initial LIGO, 
and has persisted on into the Advanced LIGO generation. 
The absence of methods to easily identify blip glitches in Initial LIGO made it
difficult to obtain meaningful statistics on the rate of such noise transients.
Furthermore, the few blips observed did not show significant correlations with any
auxiliary channels that could reveal the source of the noise. 

Two main effects complicate source identification and mitigation of blip glitches. 
First, many different physical mechanisms could be responsible for these short impulses,
and sensors may not witness these processes or record them at high enough bandwidth.
Second, the servo-control loops of the interferometer make it difficult to identify
the original sources because loop actuators may compensate for the effect of the 
original transient, propagating the impact into the main calibrated gravitational-wave channel.
In the following sections, we develop a method based on gravitational-wave searches to identify blip glitches
in \aligo{} data and study the origin of these noise transients.

\section{Identifying blip glitches in LIGO data} \label{sec:finding_blips}

Blip glitches always trigger the same region of parameter space in the \pycbc{} 
search~\cite{Canton:2014ena, Usman:2015kfa, PyCBCgithub} for gravitational waves from
compact binary coalescences. Investigation of the loudest single-detector background events
revealed that this region corresponds to short waveforms with 
total masses greater than $\simeq 75 M_{\odot}$~\cite{Nitz:2017lco,O1_DQ}.
While these waveform templates are potential blip-glitch finders, gravitational-wave searches use
various ranking statistics to separate noise transients from astrophysical events.
Particularly, the \pycbc{} search uses a high frequency sine-Gaussian $\chi^2$ 
discriminator~\cite{Nitz:2017lco} to reject a type of blip glitch that contains excess power 
at higher frequencies than expected for the waveform templates. Hence, the results
from a gravitational-wave search are not optimal to find the maximum possible number of blip glitches.
We can, however, create a blip-glitch search using a subset of short waveform templates and \pycbc{} matched-filtering techniques.
In the O1 data set, we used a reduced template bank containing 14 templates, instead of the 
250,000 used for gravitational-wave searches~\cite{PhysRevD.93.122003}.
The \pycbc{} template bank for O2 was increased in parameter space by approximately 
60$\%$~\cite{DalCanton:2017ala}, so the reduced template bank for the blip search was
increased to 30 templates. Furthermore, with the development of
\pycbc{}~Live~\cite{Nitz:2018rgo}, the O2 blip-glitch search was performed in 
low latency to rapidly diagnose variations in the rate of blip glitches.

\begin{figure}[tb]
\centering
\includegraphics[width=0.45\columnwidth]{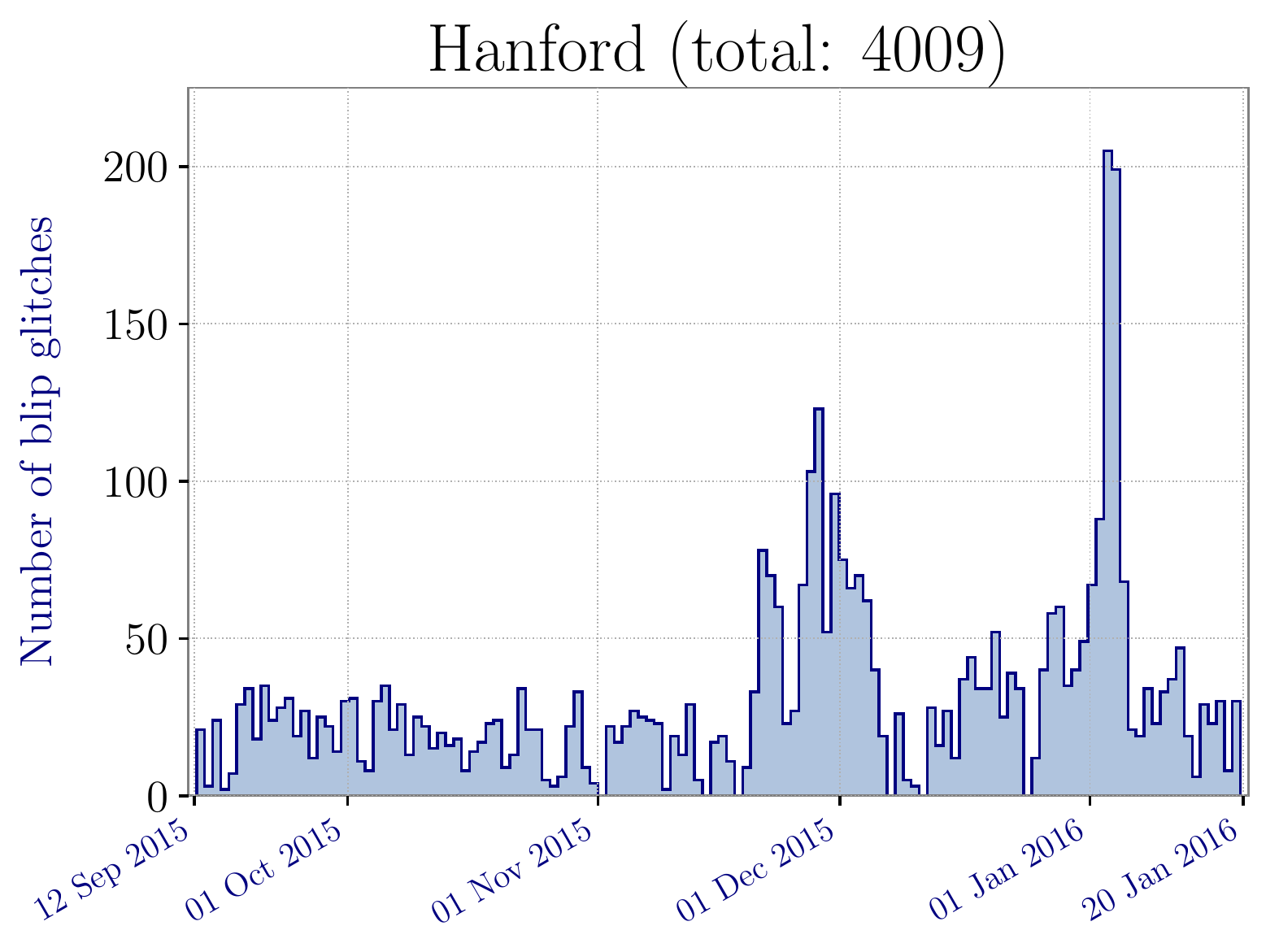}
\includegraphics[width=0.45\columnwidth]{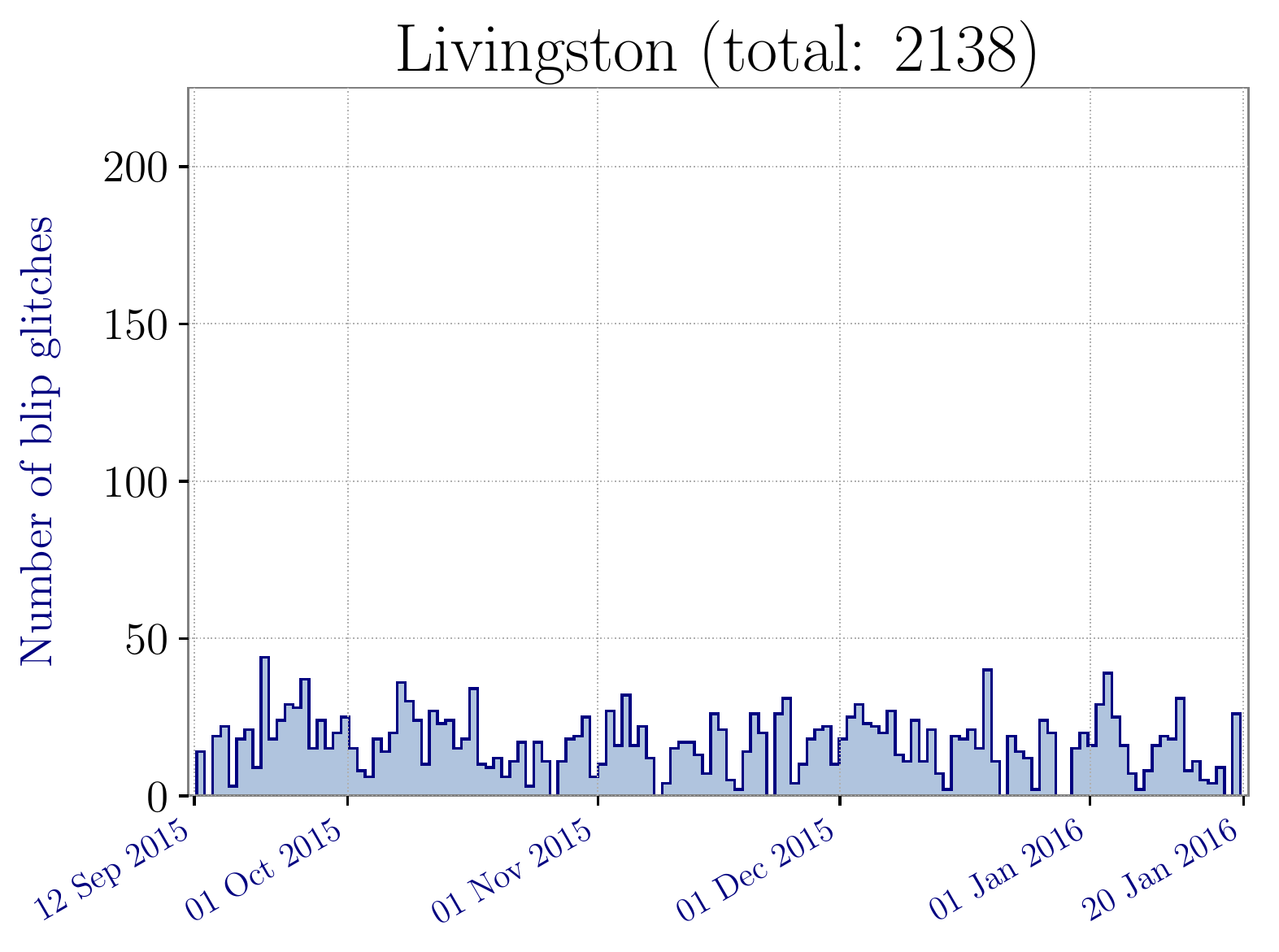} \\
\includegraphics[width=0.45\columnwidth]{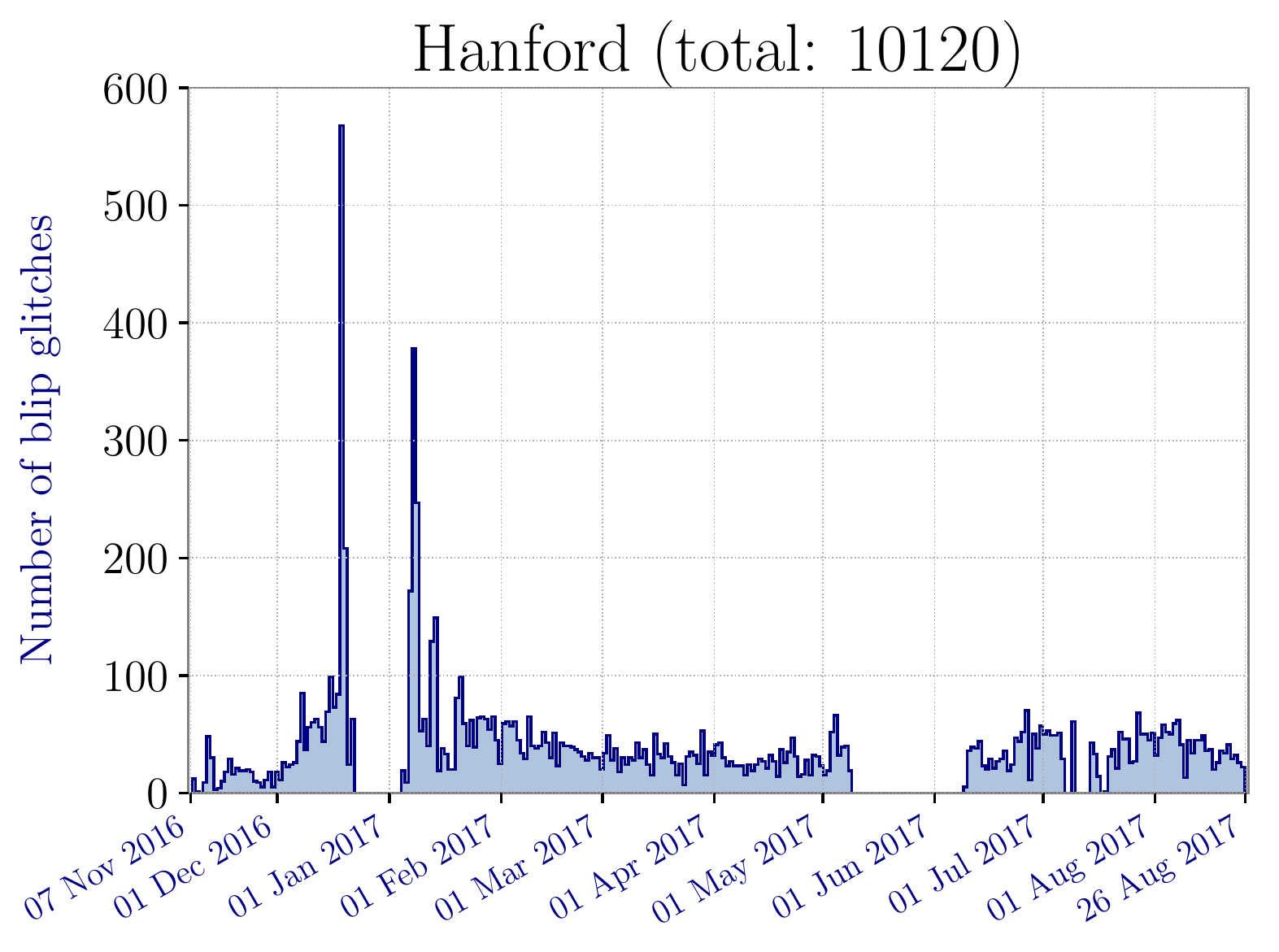}
\includegraphics[width=0.45\columnwidth]{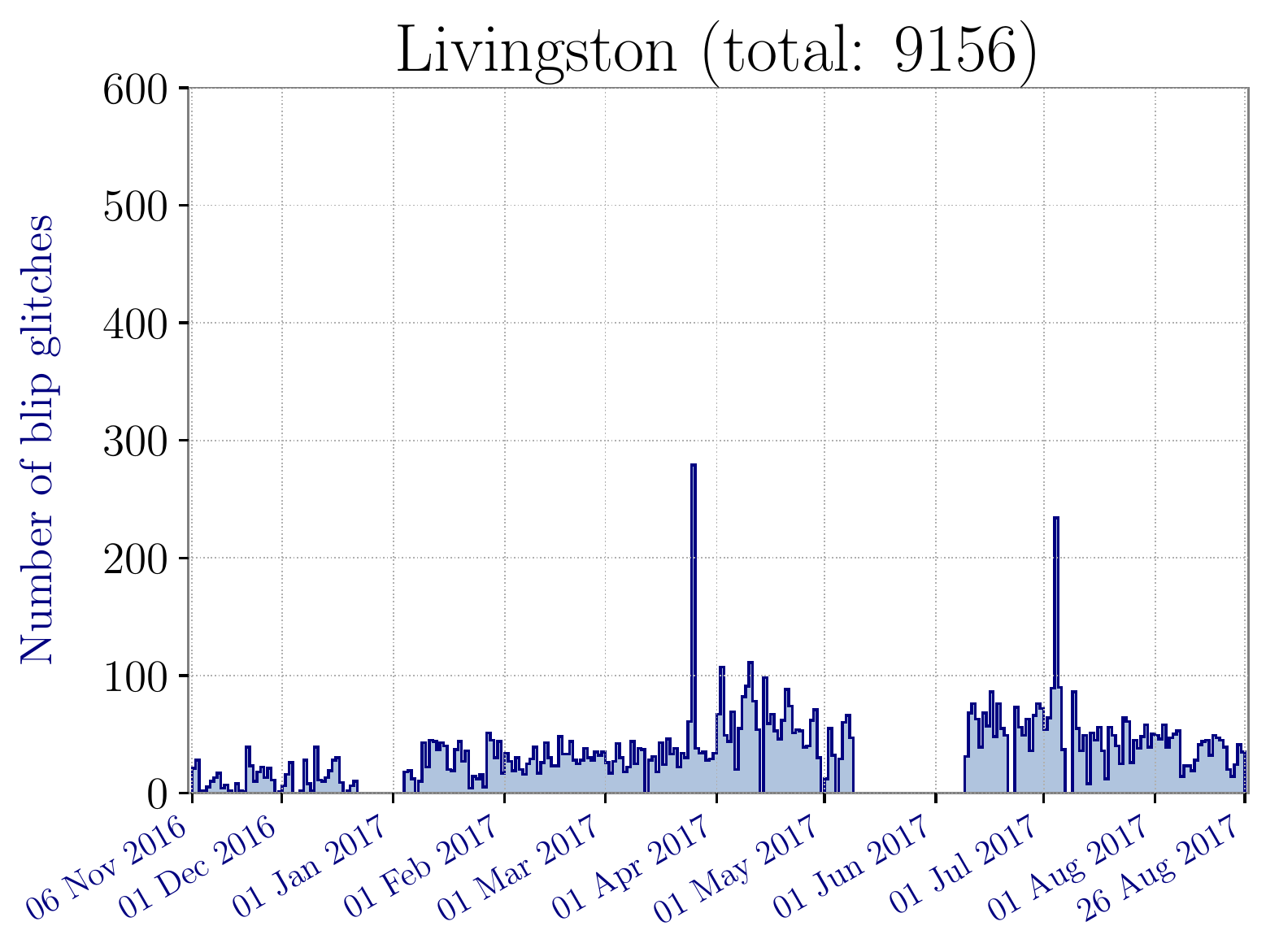}
\caption{\label{fig:hist_blips} Blip glitches found per calendar day in \aligo{} data in the four months of O1 (\textit{top row}), and the ten months of O2 (\textit{bottom row}). The left column is the Hanford observatory and the right column is the Livingston observatory. The two gaps in O2  correspond to the 2016 end-of-year holidays break and a May commissioning break to improve instruments' sensitivity.}
\end{figure}

To optimise the blip-glitch search, we define two signal-to-noise ratio (SNR) thresholds.
A lower SNR threshold of $\rho \geq 7.5$ is used to avoid random fluctuations of 
Gaussian data. An upper SNR threshold of $\rho \leq 150$ is used to avoid loud noise events
that are clearly not blips but contain similar morphology near loud, 
long-duration noise transients. 
Each single blip glitch can produce multiple search triggers, so we cluster triggers in 0.1~s time
windows after the single detector matched filtering to identify individual glitches. 
Finally, glitches that do not match our definition of a blip glitch are manually removed
(for instance, short noise transients with a small frequency bandwidth,
such as small 60 Hz power glitches). 
The resulting number of blip glitches per calendar day found in each LIGO detector 
is shown in Fig.~\ref{fig:hist_blips}, with the O1 data set in the top row and the 
O2 data set in the bottom row.

An alternative approach to identify blip glitches, not available during the 
development of the methods used here, is a new citizen science effort called 
\texttt{Gravity~Spy}~\cite{GravitySpy}. This project combines crowdsourcing with 
machine learning to categorise different types of glitches found in LIGO data. 
The glitches classified as blips in this manuscript encompass four different 
categories in the \texttt{Gravity~Spy} classification: 
blips, repeated blips, tomtes and koi fish. 
Further investigations on the different types of glitches from
\texttt{Gravity~Spy} are beyond the scope of this paper.
Here, we restrict ourselves to the blip glitches obtained with \pycbc{}.

The rate of blip glitches in each LIGO detector is given by the ratio of the number of blips 
to the duration of analysed data (analysed time), 
which is different than the calendar time used in Fig.~\ref{fig:hist_blips}. 
In the O1 run, the median blip-glitch rate in the Hanford observatory (LHO) was slightly higher than 
in the Livingston observatory (LLO), with approximately 39 blips per day of LHO analysed time 
versus 31 blips per day of LLO analysed time.
In the O2 run, the median blip-glitch rate increased in both detectors: 
approximately 47 blips per day of LHO analysed time and 48 blips per day of LLO analysed time.
The rate increase was more significant in the Livingston observatory, and both detectors
had comparable rates.

To show that blip glitches are not correlated in time between detectors, 
we perform a simple test to determine if the number of blips occurring 
within a small time difference between detectors is what would be expected 
from random coincidences.
Assuming that blip glitches are uncorrelated between both detectors, 
the expected number of coincidences within a particular time window is
\begin{equation}
\label{eq:corr_test}
\langle N_{\tau} \rangle = 2 \, | \tau | * \frac{N_{\mathrm{LHO}} * N_{\mathrm{LLO}}}{T_{\mathrm{coinc}}} \, ,
\end{equation}
with $\tau$ the maximum time difference allowed between detectors,
$T_{\mathrm{coinc}}$ the duration of coincident data (given by times 
during which both LIGO detectors were collecting data suitable for analysis),
and $N_{\mathrm{LHO}}$ and $N_{\mathrm{LLO}}$ the number of blip glitches
occurring during coincident data for Hanford and Livingston, respectively.
Table~\ref{tab:corr_test} shows the expected coincidences for different time
windows, as well as the found coincidences in the O1 and O2 data sets.
For an astrophysical signal, the maximum arrival time difference between the
two LIGO detectors is $\pm 10$ ms. In the \pycbc{} search for gravitational
waves, this time difference is allowed to be $\pm 15$ ms to account for 
timing uncertainties. As can be seen in Table~\ref{tab:corr_test}, 
there are no coincidences found within this time window. Coincidences found 
in larger time windows are consistent with the expected values.

\begin{table}[tb]
\centering
\begin{tabular}{| l || c | c | c | c |}
\hline
Maximum time difference $\tau$ (s) & $\pm$ 0.015 & $\pm$ 0.1 & $\pm$ 1 & $\pm$ 10 \\
\hline
O1 Expected $\langle N_{\tau} \rangle$ & 0.03 & 0.20 & 1.97 & 19.70 \\
O1 Found coincidences & 0 & 0 & 2 & 24 \\
\hline
O2 Expected $\langle N_{\tau} \rangle$ & 0.13 & 0.84 & 8.38 & 83.76 \\
O2 Found coincidences & 0 & 1 & 4 & 65 \\
\hline
\end{tabular}
\caption{\label{tab:corr_test} Time correlation of blip glitches between both LIGO detectors. The first row is the maximum time difference allowed between the detectors. The following rows indicate the expected number of coincidences for each time window, $\langle N_{\tau} \rangle$, and the number of coincidences found. No coincidences are found within the time window corresponding to astrophysical signals (15 ms). The number of coincidences found in larger time windows is consistent with the expected values.}
\end{table}


\section{Investigating the origin of blip glitches}
\label{sec:investigating_blips}

The fact that blips are not correlated in time between different detectors
indicates that they are likely of instrumental or environmental origin rather
than of astrophysical origin. The cosmic ray detector located at the Hanford
site has shown that there is no unusual cosmic-ray activity at times of blip
glitches~\cite{alogRomaCosmic}, and for a long time no correlations with other
environmental sensors have been detected. 
The blip glitches found with \pycbc{} tools provide, for the first time,
meaningful statistics to investigate the source of blip glitches and explore
correlations between the rate of blip glitches and the status of the detectors.
In this section we briefly explain conclusions derived from investigating
sources of blip glitches in the \aligo{} detectors. As we will see, 
these investigations indicate that there is not one single source for all blip
glitches. Instead, we find that there are different subsets of blips. 
This is not surprising, since the simple morphology of such short impulses could
be produced by many possible mechanisms without any noticeable distinction.

The LIGO detectors have radio-frequency control systems with some parts including 
analog components that interferometrically determine and control the length and
alignment of the resonant optical cavities. Some of those control systems' error and
control signals are stored digitally, but none are sampled at a rate higher than 
16384 Hz, and thus information about their signal content above $\sim$8 kHz is not
recorded. Similarly, the observatories also contain many slow servo systems and 
monitor signals which are only recorded at a rate much slower than the detector
bandwidth, i.e. 16 Hz. These systems, however, may influence and have a coupling
mechanism to the detector's primary signal at higher frequencies.  
This study has been performed using digitally stored auxiliary channels, which might not
be effective witnesses of blip glitches and their full signal content. 
Thus, even when finding hints to the origin of a subset of blip glitches, we may not be
able to fully characterise and reveal mechanisms causing these noise transients.

As a first step, we systematically looked for correlations between blip glitches and auxiliary channels
using the Hierarchical Veto (HVeto) pipeline~\cite{hveto}. This algorithm is designed to find
time correlations between transients in the main calibrated gravitational-wave channel
and in auxiliary channels with negligible sensitivity to gravitational waves (known as ``safe'' channels).
The HVeto pipeline runs daily on all glitches for detector characterisation purposes. However,
running on longer periods of data becomes computationally challenging given the large amount of
triggers produced during an observing run. With the lists of blip glitches, which constitute only a fraction of the 
totality of the triggers, we can drastically reduce the amount of data to analyse and run HVeto
on longer data segments. We found two subsets of blip glitches that showed correlations with 
auxiliary channels, one which had already been identified during the course of the run 
(described in Sec.~\ref{sec:blips_laser}) and one new subset (described in Sec.~\ref{sec:LSC-POP}).
However, the majority of blip glitches did not show correlations with the safe, fast-recording channels 
used in HVeto. Additional investigations revealed two more subsets of blip glitches, described in
Secs.~\ref{sec:blips_humidity} and~\ref{sec:computer}. 

\begin{figure}[t]
\centering
\includegraphics[width=0.49\columnwidth]{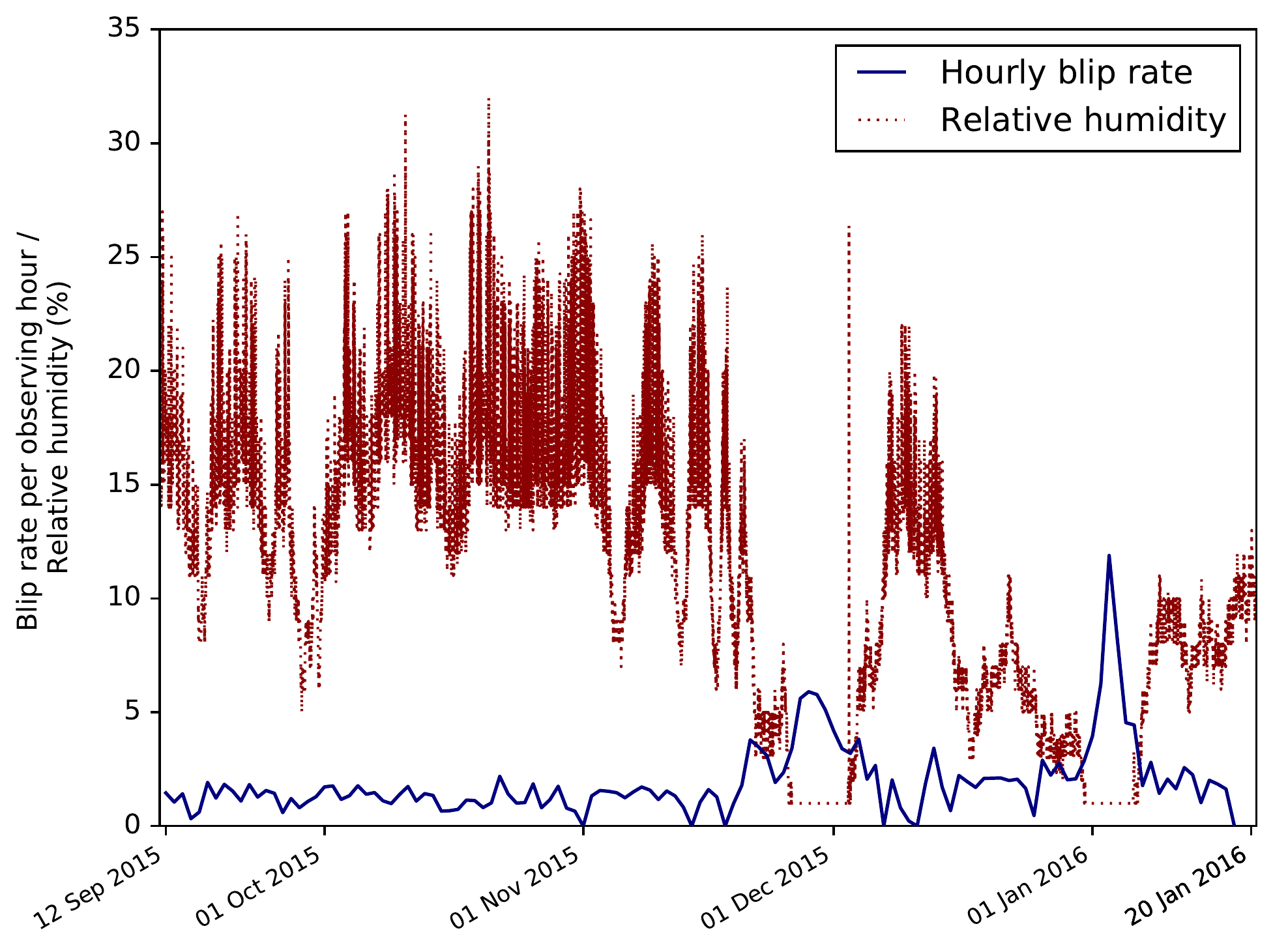}
\includegraphics[width=0.49\columnwidth]{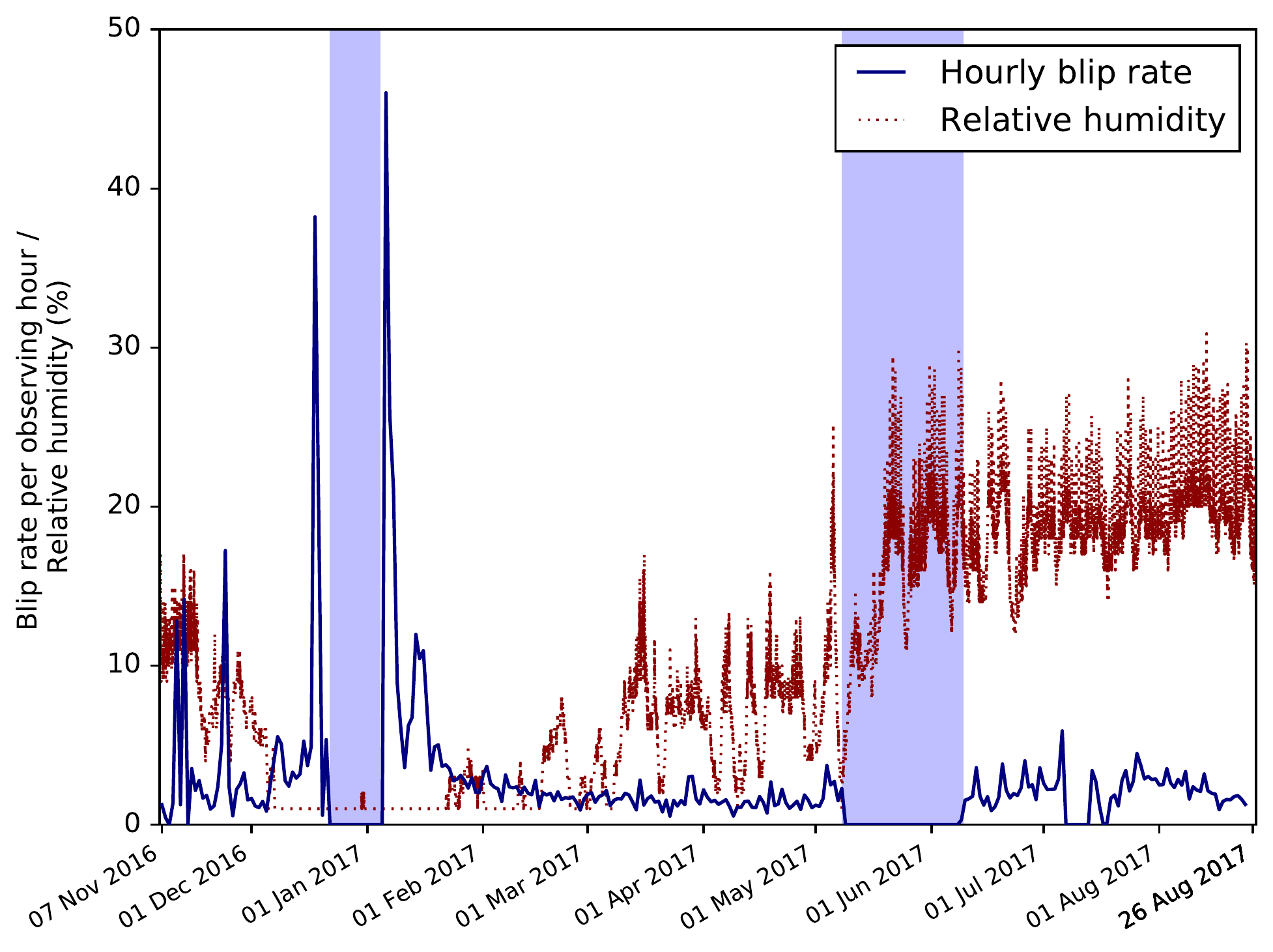}
\caption[]{\label{fig:humidity} Correlation between blip glitches and low inside relative humidity at the Hanford detector during O1 (\textit{left}) and O2 (\textit{right}). The $y$-axis shows
\begin{inparaenum}[(i)]
\item the relative humidity (RH) sensor read-back in one room of the detector's corner station (red dotted curve), and 
\item the average rate of blip glitches per hour of data (blue solid curve). Shaded regions in the O2 data correspond to the end-of-year break and the May commissioning break.
\end{inparaenum}}
\end{figure}

\subsection{Correlation with humidity} \label{sec:blips_humidity}

At the end of O1, a strong correlation between the rate of blip glitches and the outside 
temperature was found at the LIGO Hanford detector~\cite{alogSchaleHumidity}. 
The outside temperature is well correlated with the inside relative humidity: 
during the winter, heating is required to maintain an appropriate constant inside 
temperature and the inside relative humidity decreases. As can be seen in
Fig.~\ref{fig:humidity}, there was a significant increase in the rate of blip glitches
during periods when the measured relative humidity inside the building dropped below $5\%$
at the Hanford site. 
Extended periods of low relative humidity in the winter also showed an increase in the
rate of blip glitches in O2, as shown in Fig.~\ref{fig:humidity}.

The exact cause of this correlation is still not clear. Dry conditions might favour the
build up and discharge of static electricity on electronic cooling fans. Alternatively, there may be
current leakage paths that discharge in bursts when the pathways are dry. 
Therefore, it might be possible that these blip glitches are due to electronic discharges,
which could be mitigated or reduced by maintaining the inside relative humidity above critical levels. 

\subsection{Correlation with laser intensity stabilisation} \label{sec:blips_laser}

During the first half of O2, we found that a subset of blip glitches at Hanford originated
from the Pre-Stabilised Laser (PSL). A louder blip glitch with similar morphology was recorded 
at a PSL auxiliary channel at the same time as these blip glitches, indicating a strong correlation 
between the PSL and the blip in the main calibrated gravitational-wave channel.
The auxiliary channel in question witnesses one of the photodiodes at the inner loop 
of the Intensity Stabilisation Servo (ISS). This ISS is a feedback control system designed
to stabilise the power of the PSL. We briefly summarise below the parts of the PSL and the
ISS that are relevant for this section; a complete description of these systems can be 
found in~\cite{PSLfinaldesign, Kwee:12}. 

\begin{figure}[tb]
\centering
\includegraphics[width=\columnwidth]{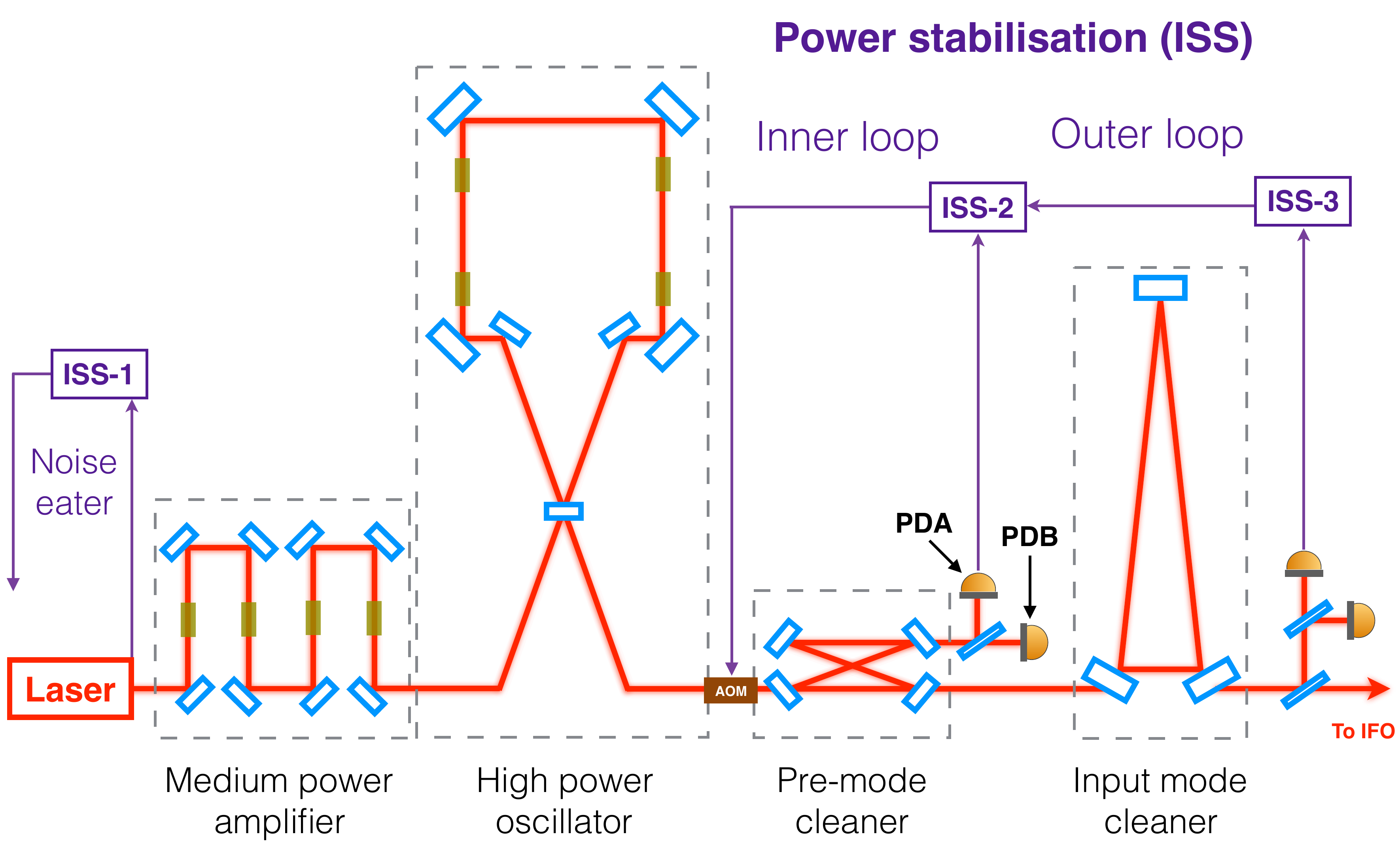}
\caption{\label{fig:PSLdesign} Simplified optical configuration of the Pre-Stabilised Laser (PSL) at the Advanced LIGO detectors (original image in~\cite{Cabero:thesis}). The input mode cleaner is not part of the PSL, but closely related. AOM is the acousto-optic modulator used as the actuator in the ISS loops. The signal recorded at the photodiode PDA witnessed the subset of PSL blip glitches occurring during the first half of O2. A similar signal was also seen at the photodiode PDB.}
\end{figure}

A simplified sketch of the optical configuration of the PSL, including all stages of 
the ISS, is shown in Fig.~\ref{fig:PSLdesign}.
The ISS consists of three loops: the noise eater, the inner loop, and the outer loop. 
The noise eater is placed directly at the laser output to reduce the relaxation oscillation of the laser. 
The inner and outer loops are situated after the Pre-Mode Cleaner (PMC) to stabilise the
power in the detection frequency band (10 Hz to 10 kHz).
The PMC, which improves the quality and pointing stability of the laser beam, 
passively suppresses power fluctuations at radio frequencies only.
Power fluctuations at lower frequencies are measured downstream of the PMC by
one pick-off port of the PMC (the photodiode labelled PDA in Fig.~\ref{fig:PSLdesign}).
Through an analog electronics feedback control loop and an acousto-optic modulator (AOM)
as actuator, the inner loop stabilises these lower-frequency fluctuations. 
The outer loop is responsible for the ultimate power stability in the interferometer.
This loop, implemented to achieve the required power stability at 10 Hz, 
senses the light from a photodetector directly upstream of the interferometer. 
Furthermore, the outer loop improves the power stability of the inner loop and 
compensates for power noise that is not suppressed in the previous stages. 

The PSL blip glitches occurred from the beginning of the O2 run (November 2016) 
until the end of February 2017. A total of 25 blip glitches correlated to the ISS 
were found in that period, approximately 0.25\% of the population of blip glitches 
reported in Sec.~\ref{sec:finding_blips} for the entire O2 data set of the LIGO Hanford observatory. 
The correlation with the auxiliary channel recording the signal from the PDA allowed
to identify instances of PSL blip glitches using Omicron\footnote{
Omicron~\cite{Omicron} is an algorithm used to identify excess power in gravitational-wave
data via a constant-$Q$ time-frequency wavelet transform~\cite{Qtransform}. 
This algorithm is used to find noise transients in the gravitational-wave channel 
and in the auxiliary channels.}.
These glitches were then vetoed in the gravitational-wave searches.
The exact cause of the laser glitching remains unknown.

\subsection{Correlation with computer errors} \label{sec:computer}

The Advanced LIGO detectors have a complex data transfer and acquisition architecture. 
Many subsystems and channels require real-time digital controls, 
and real-time analog data acquisition for analysis and archiving~\cite{AdvLIGO}. 
Data acquisition and control applications for the various subsystems are defined 
and built using the Real-Time Code Generator (RCG)~\cite{RCG}. The RCG control systems describe the
desired execution sequence for different code infrastructures.

Interferometer sensor and actuator signals flow between Analog-to-Digital and 
Digital-to-Analog Converters (ADC and DAC) through a customised computer infrastructure,
located in rooms separate from the interferometer instrumentation. 
The global control system is necessarily distributed between the corner station and
end-station buildings housing the detectors. Thus, the computers that sense and control
the interferometer are in several, often distant, physical locations. They are digitally
connected via commercial, computer-to-computer, high-speed network hardware and optical
fibre. This Inter-Process Communication (IPC), both between individual cores of computers
and between computers, could have errors that cause blip glitches.

The suspension (SUS) system is the actuator of the global length and angular control 
system, and thus the primary control loop controlling the differential arm lengths. 
Between O1 and O2, SUS control systems became too complicated for the standard real-time 
(front end) computers. Faster computers were installed at both LIGO sites to speed
up both the SUS and the I/O Processor (IOP) control systems. The IOP is in charge of
interfacing with the I/O hardware modules, and of synchronising with the interferometer 
timing system~\cite{AdvLIGO}. 
During the course of O2, a subset of blip glitches was identified in coincidence with
times of computer errors related to the IOP and SUS control systems. 
While there are several different types of computer errors, we restrict ourselves here 
to the errors that were observed in coincidence with blip glitches during O2.

Timing errors appear, for instance, when the time between code cycles is outside the
assigned limits, or when the code execution time is greater than the allowed time.
An ADC overrun error can originate from:
\begin{inparaenum}[(i)]
\item data not arriving on time from the ADC modules, or
\item channel misidentification by the IOP control system, which reads the ADC signal 
at the beginning of each cycle and assigns it to the corresponding channel.
\end{inparaenum}
The ADCs have channel identification, and can therefore re-synchronise themselves 
when the error is associated to the latter. In that case, the RCG discards the data from
the corresponding corrupted cycle. However, if the ADC error is associated to a timing
error, it can propagate through the IPC to downstream control systems. 
Finally, an IPC error can indicate a fault in receiving IPC data via any IPC mechanism~\cite{RCG}. 
Computer errors seem to happen more frequently in Hanford than in Livingston, 
even though the two systems are identical.

\begin{table}[tb]
\centering
\begin{tabular}{| l | c | c |}
\hline
Control system & Total errors & Isolated errors \\
\hline
(1) IOP-SUSE\{X/Y\} & 1778 & - \\
(2) SUS-ETM\{X/Y\} & 1313 & 8 \\
(3) SUS-TMS\{X/Y\} & 129 & 1 \\
(4) SUS-ETM\{X/Y\}PI & 889 & 7 \\
(5) IOP-SEIE\{X/Y\} & 1732 & 13 \\
(6) ALS-E\{X/Y\} & 1180 & 7 \\
(7) ISC-E\{X/Y\} & 1197 & 7 \\
\hline
\end{tabular}
\caption{\label{tab:errors} All the end station X,Y (E\{X/Y\}) control systems we look at to find computer errors: (1) IOP Suspension, (2) quad Suspension, (3) Transmission Suspension, (4) Parametric Instability, (5) IOP Seismic Isolation, (6) Arm Length Stabilisation, (7) Interferometer Sensing and Controls. The second column indicates the total number of errors in that system. The third column indicates how many of the errors in that control system did not propagate from the IOP-SUS system.}
\end{table}

It is still uncertain how the computer errors are causing a glitch in the main calibrated 
strain channel. One possibility could be the corrupted signal from the SUS getting to the 
interferometer sensing control (ISC) system. However, there
does not seem to be a clear correlation between the type of error at the times of blip 
glitches: we have observed timing as well as ADC and IPC errors.
There are auxiliary channels for each control system that record occurrences of any
type of computer errors. Here we look at seven systems between the corner station and both
arm ends of the interferometer. 
In O2 data, the system with the largest number of computer errors is the IOP-SUS system, 
with 1778 errors on both end-X and end-Y (E\{X/Y\}) systems (832 and 946, respectively). 
From the blip glitches with SNR $\rho < 150$ identified in Sec.~\ref{sec:finding_blips}, 
a total of 625 blips are in coincidence with IOP-SUS front-end computer errors.  
Computer errors also induce glitches louder than $\rho = 150$, which are not included in 
our lists of blip glitches. Hence, the percentage of IOP-SUS front-end computer errors 
associated with glitches in the main calibrated strain channel is currently unknown, 
and we can only estimate it to be larger than 35\%.

Errors in the IOP-SUS system can propagate to downstream control systems. 
In Table~\ref{tab:errors} we indicate the number of errors found in the different systems,
and how many of those errors did not propagate from the IOP-SUS system (isolated errors). 
Since there are some coincidences between the downstream systems, 
only 14 of the isolated errors are new with respect to the IOP-SUS system, 
6 of which are in coincidence with a blip glitch. 
Hence, we find 631 blip glitches with SNR $\rho < 150$ correlated with computer errors, 
approximately 6.2\% of the population of blip glitches reported in Sec.~\ref{sec:finding_blips}
for the O2 data of the LIGO Hanford observatory. 

\subsection{Correlation with power recycling cavity control signals} \label{sec:LSC-POP}

\begin{figure}[tb]
\centering
\includegraphics[width=0.9\columnwidth]{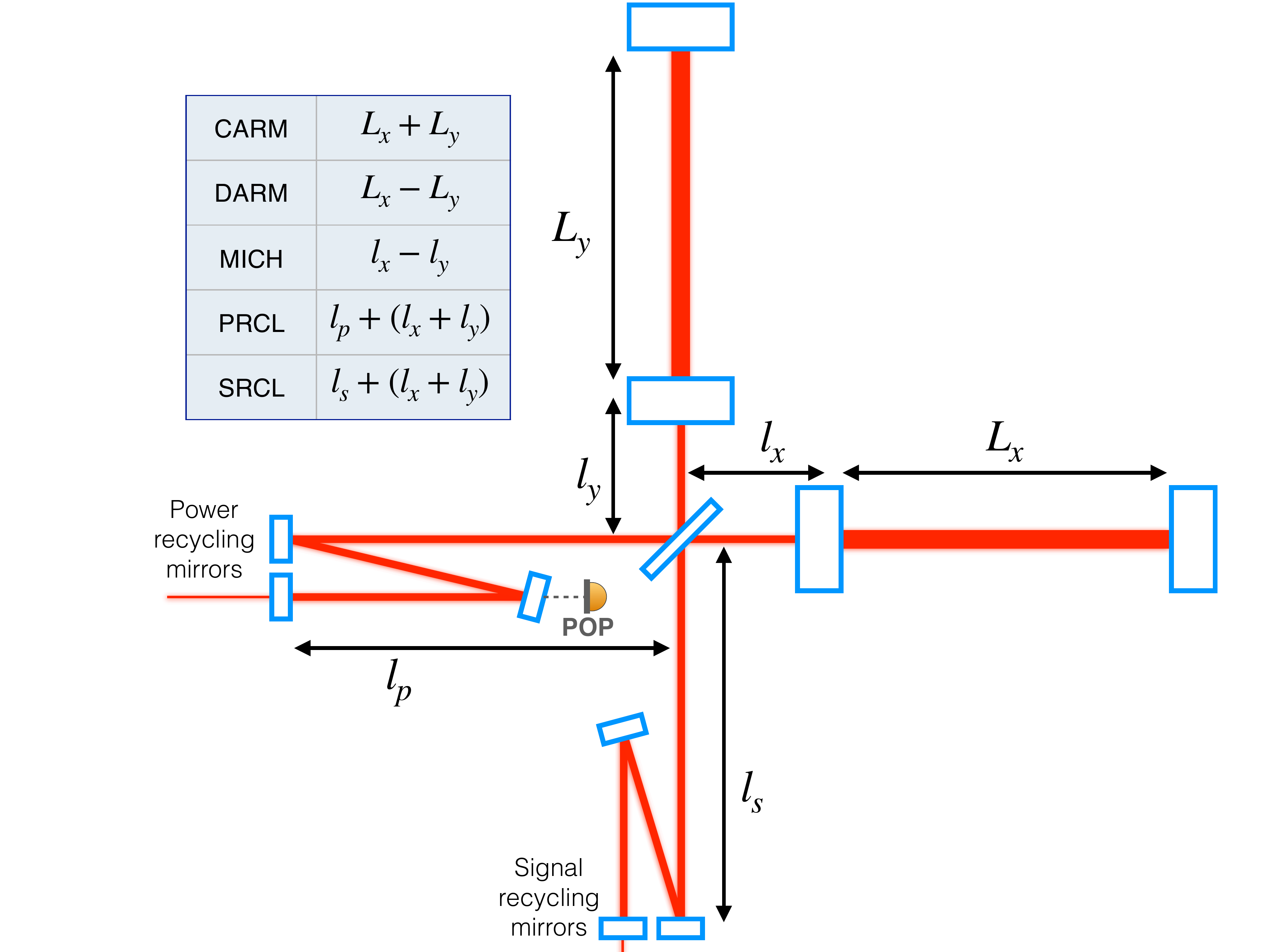}
\caption{\label{fig:DoF} Simplified configuration of the five primary resonant cavities in the Advanced LIGO detectors: the Common ARM cavity length (CARM), the Differential ARM cavity length (DARM), the MICHelson length (MICH), the Power Recycling Cavity Length (PRCL), and the Signal Recycling Cavity Length (SRCL). 
The Pick-Off port for the Power recycling cavity (POP) witnessed a subset of blip glitches occurring in O2.
}
\end{figure}

Figure~\ref{fig:DoF} shows the five primary resonant cavity systems in the Advanced LIGO
detectors\footnote{There are different conventions in the literature describing the cavity
basis. Some include a factor of two~\cite{izumi2016advanced} while others do 
not~\cite{PhysRevLett.116.131103}.}. 
The gravitational-wave channel is related to the reconstruction of differential length
variation between the two arm cavities,
$L_x - L_y$ (with $L_x$ and $L_y$ the length of the X~arm and the Y~arm, respectively). 
However, in order for the interferometer output to remain linearly proportional 
to $L_x - L_y$, all five cavity systems must be controlled.

The cavities are controlled via an interferometric, Pound-Drever-Hall~\cite{drever1983laser}, 
radio-frequency (RF) demodulation scheme, described briefly as follows. 
Phase-modulated sidebands are intentionally instilled on the laser light input into the 
interferometer via electro-optic modulators. After resonating inside the interferometer's 
optical cavities, this light is then captured on photodiodes positioned at various 
strategic pick-off ports. Thus, the pick-off port signal is a probe of the length
(and alignment) of the cavity, serving as an error signal for the cavity length 
(or alignment) control system. The raw photodiode current signal is demodulated at the
sideband frequencies and conditioned via a system of analog electronics. The resulting
conditioned, demodulated signal is passed into either analog or digital control servos,
depending on the necessary bandwidth for control. 
See further details of the control system in~\cite{izumi2016advanced}. 
As a part of the out-of-loop diagnostics for this system, the raw light power levels 
on the photodiodes are also digitised and stored.
	
As mentioned at the beginning of Sec.~\ref{sec:investigating_blips}, it is impractical 
to digitise all signals of this RF electronics system with high bandwidth. 
Therefore, this study can only use those signals that are stored to suggest correlations 
between the control system and the gravitational wave channel. 
In O2, we found that the diagnostic power level on one of these RF photodiodes 
-- namely the Pick-Off port for the Power recycling cavity (POP) -- 
showed loud blip glitches at the same time as a subset of blip glitches in the main 
calibrated channel, with comparable morphologies in both channels. 
This correlation has been seen in both LIGO detectors: 168 blips in O2 Hanford data 
and 199 blips in O2 Livingston data showed correlation with the POP diagnostic channel. 
That makes approximately 1.7\% of the O2 population of blip glitches reported in 
Sec.~\ref{sec:finding_blips} for the LHO data, and 2.2\% of the O2 population for 
LLO data. Further investigations to understand the source of these glitches and possibly
mitigate them are currently ongoing.

\section{Conclusions} \label{sec:conclusions}

In this paper we described a type of noise transients in gravitational-wave data 
commonly known as blip glitches. 
During the first two observing runs of Advanced LIGO, thousands of blips were found 
in each LIGO detector using \pycbc{} techniques. 
Blip glitches significantly reduce the sensitivity of searches for high mass compact binaries. 
It is therefore important to identify the origin of these noise events so they can be 
mitigated from the searches. With the lists of blip glitches obtained, we conducted 
investigations on the origin of blip glitches and reported correlations with 
four different source channels: low relative humidity inside the building, 
laser intensity stabilisation, computer errors, and power recycling control signals.

Despite the importance of identifying sources of blip glitches, these were only four small 
subsets of all the blip glitches found. In total, less than 8\% of LHO blip glitches and 
about 2\% of LLO blip glitches in O2 data have shown a correlation with an auxiliary channel.
Search pipelines are evolving to new ranking statistics that veto particular types of blip 
glitches from the searches (see for instance~\cite{Nitz:2017lco}). 
At the same time, investigations at the LIGO detectors continue during commissioning and 
observing runs. Now that gravitational waves from coalescing black holes represent an 
important astrophysical output of LIGO and Virgo, it becomes even more urgent to mitigate 
blip glitches from the data. Otherwise, we might not be able to distinguish 
between blip glitches and marginal gravitational-wave signals from high mass systems.

\section{Acknowledgements}

We would like to thank Jenne Driggers, Sheila Dwyer, Michael Landry and Jessica McIver 
for interesting discussions. We also thank the LIGO Lab and the whole team at the LIGO 
Hanford observatory for investigations performed at the site.
MC acknowledges support from NSF grant PHY-1607449, the Simons Foundation, and the Canadian Institute For Advanced Research (CIFAR). 
TD acknowledges support from the Maria de Maeztu Unit of Excellence MDM-2016-0692. 
LKN received funding from the European Union Horizon 2020 research and innovation programme 
under the Marie Sklodowska-Curie grant agreement No 663830. 
This paper uses data from the Advanced LIGO detectors and computations have been performed 
on the LIGO Data Grid and on the Atlas computer cluster at the Albert Einstein Institute (Hannover).

\section*{References}

\bibliography{Bibliography}

\end{document}